\documentclass[onecolumn,aps,noshowpacs,11pt]{revtex4}

\usepackage{amsmath,amsfonts,amsthm,amssymb,latexsym,color,palatino}
\usepackage[colorlinks=true,urlcolor=webblue,linkcolor=webgreen,filecolor=webblue,citecolor=webgreen,pdfpagemode=UseOutlines,pdfstartview=FitH,pdfpagelayout=OneColumn,bookmarks=true]{hyperref}

\hypersetup{
  pdftitle=Strong Fourier Sampling Fails over $G^n$,
  pdfauthor=Gorjan Alagic and Cristopher Moore and Alexander Russell}

\def\identity{\leavevmode\hbox{\small1\kern-3.2pt\normalsize1}}
\def\bra{\left\langle}
\def\ket{\right\rangle}
\def\ZZ{\mathbb Z}
\definecolor{webgreen}{rgb}{0,.5,0}
\definecolor{webblue}{rgb}{0,0,.5}

\numberwithin{equation}{section}

\newtheorem{theorem}{Theorem}
\newtheorem{lemma}{Lemma}
\newtheorem{prop}{Proposition}

\begin{document}

\title{Strong Fourier Sampling Fails over $G^n$}

\author{Gorjan Alagic$^1$}
\email{alagic@math.uconn.edu}
\author{Cristopher Moore$^2$}
\email{moore@cs.unm.edu}
\author{Alexander Russell$^3$}
\email{acr@cse.uconn.edu}
\affiliation{
$^1$Department of Mathematics, University of Connecticut\\
$^2$Department of Computer Science, University of New Mexico\\
$^3$Department of Computer Science and Engineering, University of Connecticut}

\date{\today}

\begin{abstract}
  We present a negative result regarding the hidden subgroup problem
  on the powers $G^n$ of a fixed group $G$. Under a condition on the
  base group $G$, we prove that strong Fourier sampling cannot
  distinguish some subgroups of $G^n$. Since strong sampling is in
  fact the optimal measurement on a coset state, this shows that we
  have no hope of efficiently solving the hidden subgroup problem over
  these groups with separable measurements on coset states (that is,
  using any polynomial number of single-register coset state
  experiments). Base groups satisfying our condition include all
  nonabelian simple groups. We apply our results to show that there
  exist uniform families of nilpotent groups whose normal series factors
  have constant size and yet are immune to strong Fourier sampling.
\end{abstract}

\maketitle

\section{Introduction}

\subsection{Results}

In this article, we consider the problem of determining hidden
subgroups in the powers $G^n$ of a fixed base group $G$. We first
prove a preliminary result, which assumes only that the base group $G$
contains an involution $\mu$ which is not an element of the center
$Z(G)$.  Under this assumption, we prove that weak Fourier sampling
cannot distinguish the trivial subgroup from the subgroup $H = \{1,
m\}$, where $m$ is chosen uniformly at random from the conjugacy class
$[(\mu,\dots,\mu) ]$. The condition $\mu \notin Z(G)$ is equivalent to the existence of an
irreducible representation $\rho$ of $G$ such that
\begin{equation}\label{char-bound-condition}
\left|\chi_\rho(\mu)\right| < d_\rho.
\end{equation}

Our main result essentially requires that the dimensionwise mass of
the group algebra of $G$ consisting of irreducibles that do \emph{not}
satisfy (\ref{char-bound-condition}) is less than the expected
dimension of an irreducible chosen according to the Plancherel
distribution on $\widehat G$. Under this stronger condition, we show
that even strong Fourier sampling cannot distinguish $H$ from the
trivial subgroup. Since strong sampling is in fact
optimal~\cite{MRS05}, this shows that we cannot solve the HSP on these
groups with any polynomial number of (single-register) experiments on
coset states.

In the last section, we prove that all nonabelian simple groups
satisfy this condition, and use our results to construct solvable
groups and nilpotent groups immune to strong Fourier sampling in this
sense.  These groups are interesting because they have normal series
whose factors are of \emph{constant} size. This can be seen as an
explanation for why recursive approaches to solving the hidden
subgroup problem have been constrained to groups of such special
structure, as those handled by \cite{HallgrenRT00},
\cite{FriedlIMSS02}, or \cite{MooreRRS04}.

In the remainder of the introduction, we summarize some necessary
results and observations from~\cite{MRS05}. The second section
consists of a simple proof of the failure of weak sampling when the
base group contains an involution $\mu$ not in the center. The third
section is dedicated to proving that strong sampling fails under the
stronger condition described above. The fourth section contains some
relevant examples.

\subsection{Optimal measurements}

As discussed in~\cite{MRS05}, the optimal \emph{positive
  operator-valued measurement} (POVM) for the hidden subgroup problem
on a group $G$ consists of strong Fourier sampling. In this optimal
measurement, we first measure the representation name $\rho$, and then
perform a POVM on the space of $\rho$, whose possible outcomes are unit
vectors $B = \{{\bf b}\}$ which satisfy the completeness condition
\begin{equation}\label{completeness}
\sum_{\bf b} a_{\bf b} |{\bf b}\rangle \langle {\bf b} | = \identity.
\end{equation}
for some positive real weights $a_{\bf b}$. This condition implies
that $B$ is a possibly over-complete basis, or \emph{frame}, for $\rho$.
Applying the Fourier transform to the initial state
$$
|H \rangle = \frac{1}{\sqrt {|H|}}\sum_{h \in H} |h \rangle
$$
yields
$$
\widehat H(\rho) = \sqrt{\frac{d_\rho}{|H||G|}}\sum_{h \in H} \rho(h) =
\sqrt{\frac{d_\rho|H|}{|G|}}\Pi_H^\rho
$$
where $\Pi_H^\rho$ is the projection operator $|H|^{-1}\sum_{h \in H}\rho(h)$. The
probability that a particular $\rho$ is observed through weak Fourier
sampling is then
\begin{equation}\label{weak-samp-distrib}
  P_H(\rho) = \left \| \widehat H(\rho)\right \|^2 = \frac{d_\rho|H|}{|G|} {\bf rk}~\Pi_H^\rho.
\end{equation}
If we continue with strong sampling, i.e. measuring within the representation $\rho$,
then the conditional probability that we observe the vector ${\bf b}$, given that we have
observed $\rho$, is given by
$$
P_{H, \rho} ({\bf b}) = a_{\bf b}\frac{\left \|\widehat H(\rho){\bf b}\right \|^2}{P_H(\rho)}
= a_{\bf b}\frac{\left \|\Pi_H^\rho {\bf b}\right \|^2}{{\bf rk}~\Pi_H^\rho}.
$$
When $H$ is the trivial subgroup, $\Pi_H^\rho = \identity_{d_\rho}$ and hence
\begin{equation}\label{natural}
P_{\{1\}, \rho} = \frac{a_{\bf b}}{d_\rho}.
\end{equation}
We will refer to $P_{\{1\}, \rho}$ as the \emph{natural distribution} on $B$. Note
that the natural distribution is equal to the uniform distribution when $B$ is just
a basis.

\subsection{Previous results}

In our analysis of strong Fourier sampling on $G^n$, we will need to carefully control 
the expectation and variance of the term $\left\|\Pi_H^\rho {\bf b}\right\|^2$ 
in $P_{H, \rho}({\bf b}).$ When the representation $\rho$ in question is clear, and 
the subgroup is $\{1, m\}$, we let
$$
\Pi_m = \Pi_{\{1, m\}}^\rho = \frac{\rho(1) + \rho(m)}{2}.
$$
for $m \in G.$ Recall that any reducible representation $\tau$ of a group can be written as a direct 
sum $\tau = \bigoplus_{\sigma \prec \rho} a_\sigma \sigma$ of irreducible 
representations $\sigma$ of the same group, each with multiplicity $a_\sigma$. Let $\Pi^\tau_\sigma$ 
denote the projection operator, operating on the space of $\tau$, whose image is the 
subspace of all the irreducibles isomorphic to $\sigma$. With this notation, we have the following 
lemma from \cite{MRS05}:
\begin{lemma} \label{exp-var-lemma}
Let $\rho$ be an irreducible representation acting on a space $V$ and
let ${\bf b} \in V.$ Let $m$ be an element chosen uniformly at random from a conjugacy
class $[m]$ of involutions. Then
\begin{align*}
\text{Exp}_m \|\Pi_m{\bf b}\|^2 & = \frac{1}{2}\|{\bf b}\|^2\left(1+ \frac{\chi_\rho\left([m]\right)}{d_\rho}\right) \\
\text{Var}_m \|\Pi_m{\bf b}\|^2 & \leq \frac{1}{4}\sum_{\sigma \prec \rho \otimes \rho^*} \frac{\chi_\sigma\left([m]\right)}{d_\sigma}
\left \|\Pi^{\rho \otimes \rho^*}_\sigma \left({\bf b} \otimes {\bf b}^*\right)\right \|^2.
\end{align*}
\end{lemma}
Under the conditions of the lemma, we also have
\begin{equation}\label{expect}
\text{Exp}_m \left\|\Pi_m {\bf b}\right\|^2 =  \|{\bf b}\|^2\frac{{\bf rk}~\Pi_m}{d_\rho}
\end{equation}
which, by the lemma, yields
\begin{equation}\label{rankbound}
{\bf rk}~\Pi_m = \frac{d_\rho}{2}\left(1 + \frac{\chi_\rho(m)}{d_\rho}\right)
\end{equation}
We will also need the following fact from \cite{MRS05}:
\begin{lemma}\label{projection-lemma}
Let $B = \{{\bf b}\}$ be a collection of unit vectors in an irreducible representation $\rho$, 
satisfying the completeness condition (\ref{completeness}). Let $L$ be a subspace of $\rho \otimes \rho^*$
and $\Pi_L$ the projection operator onto $L$. Then
$$
\sum_{\bf b \in B} a_{\bf b} \left \| \Pi_L\left({\bf b} \otimes {\bf b}^*\right)\right \|^2 \leq \dim L.
$$
\end{lemma}

\section{Weak Fourier Sampling on $G^n$}

In this section, we establish two relatively easy results regarding weak Fourier sampling
on $G^n$. The first of these shows that if $G$ contains an involution which is not in the center,
then weak Fourier sampling cannot distinguish certain subgroups of $G^n$ from each other or the trivial subgroup.
For an element $g$ of the base group $G$, let $[(g, \dots, g)] \subset G^n$ denote the $G^n$-conjugacy
class of the element $(g, \dots, g)$. Notice that this is the same as the set
$[g]^n = [g] \times \cdots \times [g] \subset G^n$ formed by the $n$-fold product
of the $G$-conjugacy class $[g]$ of $g.$

\begin{lemma} \label{weak-samp-thm}
Let $G$ be a group with an involution $\mu \notin Z(G)$, and let $H = \{1, m\} \leq G^n$ 
where $m$ is chosen uniformly at random from the conjugacy class $[(\mu, \dots, \mu)]$. 
Then the total variation distance between the weak Fourier sampling distributions 
for $H$ and $\{1\}$ is at most $2^{-n/2}$.
\end{lemma}
\begin{proof}
We upper bound the total variation distance between the distributions in question:
\begin{align*}
\left \|P_{\{1\}} - P_{H} \right \|_1 
= &\sum_{\rho \in \widehat{G^n}} \left| \frac{d_\rho}{|G|^n}{\bf rk}~\Pi_{\{1\}}^\rho - \frac{2d_\rho}{|G|^n}{\bf rk}~\Pi_m \right| \\
 = &\frac{1}{|G|^n} \sum_{\rho \in \widehat{G^n}} \left| d_\rho^2 - d_\rho^2 \left(1 + \frac{\chi_\rho(m)}{d_\rho}\right) \right|
= \frac{1}{|G|^n} \sum_{\rho \in \widehat{G^n}} \left| d_\rho \cdot \chi_\rho(m)\right|
\end{align*}
where we have used equation (\ref{rankbound}) in the second step. Viewing the last line as an inner product,
we apply Cauchy-Schwarz to get
\begin{align*}
\left \|P_{\{1\}} - P_{H} \right \|_1 \leq &\frac{1}{|G|^n}\left(\sum_\rho d_\rho^2\right)^{1/2}\left(\sum_\rho \chi_\rho(m)\chi_\rho^*(m)\right)^{1/2}
= \frac{1}{|G|^{n/2}}\left(\sum_{\rho \in \widehat{G^n}} \chi_\rho(m)\chi_\rho^*(m)\right)^{1/2}\\
 = &\frac{1}{|G|^{n/2}}\left(\sum_{\rho_1, \dots, \rho_n \in \widehat{G}} \chi_{\rho_1}(\mu) \cdots \chi_{\rho_n}(\mu) \chi_{\rho_1}^*(\mu) \cdots \chi_{\rho_n}^*(\mu)\right)^{1/2} \\
 = &\frac{1}{|G|^{n/2}}\left(\sum_{\rho \in \widehat{G}} \chi_\rho(\mu) \chi_\rho^*(\mu)\right)^{n/2}
 = \left(\frac{\chi_C(\mu)}{|G|}\right)^{n/2}.
\end{align*}
where $\chi_C$ is the character of the conjugation representation of $G$. 
This is just the number of fixed points of the conjugation action of $\mu$ on $G$, 
i.e. the size of the centralizer $C_\mu$. As $\mu$ is not in the center, $C_\mu$ 
is a proper subgroup, and hence $\chi_C(\mu) \leq |G|/2$, which completes the proof.
\end{proof}

The next lemma shows that, for any nonabelian $G$, weak Fourier sampling 
the initial state corresponding to a constant-size subgroup of $G^n$ almost always results 
in a representation of very large dimension. Unless specified otherwise,
it is understood that expectations and variances are taken over representations selected 
from $\widehat G$ according to $\mathcal P_G$, the Plancherel distribution.

\begin{lemma} \label{big-irrep-lemma}
Let $G$ be a nonabelian group, and suppose $\vec \rho \in \widehat{G^n}$ is the result of weak Fourier sampling the initial
state corresponding to a subgroup of order $k$. Then, given any constant $0 < \beta < \emph{Exp}_\tau \left[\log d_\tau\right]$, 
there is a constant $\alpha > 0$ such that the dimension of $\vec \rho$ is greater than $e^{\beta n}$ with probability at 
least $1 - 2ke^{-\alpha^2 n/4}.$
\end{lemma}

\begin{proof}
Define the following constants, depending only on the base group $G$:
$$
M = \log \max\{d_\tau : \tau \in \widehat G \} \qquad \text{and} \qquad
w = \text{Exp}_\tau  \frac{\log d_\tau}{M}.
$$
Recall that the probability of measuring $\vec \rho$ through weak Fourier sampling is
$$
P_{H}(\vec \rho) = \frac{kd_{\vec \rho}}{|G^n|}{\bf rk}~\Pi_H^{\vec \rho} \leq \frac{kd_{\vec \rho}^2}{|G|^n}.
$$
This is never more than $k$ times the probability assigned to $\vec \rho$ by $\mathcal P_{G^n}$. 
If we can prove that choosing $\vec \rho$ according to $\mathcal P_{G^n}$ 
results in a large representation with probability $1 - \epsilon$, then weak sampling 
will result in such a representation with probability at least $1 - k\epsilon$.
Equivalently, we can choose $n$ representations $\rho_i$ according to $\mathcal P_G$, 
and denote the resulting representation $\rho_1 \otimes \rho_2 \otimes \cdots \otimes \rho_n$ of $G^n$ by $\vec \rho.$ 
We wish to consider the i.i.d. random variables $X_i = (\log d_{\rho_i})/M - w$,
which satisfy Exp$_{\rho_i}X_i = 0$. Since $0 < (\log d_{\rho_i})/M \leq 1$, and shifting
a random variable does not affect variance, we also have
$$
\text{Var}_{\rho_i} X_i = \text{Var}_{\rho_i} \frac{\log d_{\rho_i}}{M}
\leq \text{Exp}_{\rho_i}\left(\frac{\log d_{\rho_i}}{M}\right)^2 \leq 
\text{Exp}_{\rho_i}\frac{\log d_{\rho_i}}{M} = w.
$$
By the independence of the $X_i$, it immediately follows that
$$
\text{Exp}_{\vec \rho \in \widehat G}\sum_i X_i = 0 \qquad \text{and} \qquad 
\text{Var}_{\vec \rho \in \widehat G}\sum_i X_i\leq nw.
$$
We can now apply the following Chernoff bound:
$$
\text{Pr}\left[ \left|\sum_i X_i \right| \geq \lambda \sqrt{nw}\right] \leq 2e^{-\lambda^2/4}
\qquad \text{for} \qquad 0 \leq \lambda \leq 2\sqrt{nw}.
$$
Let us assume that the sum of the $X_i$ is less than $\lambda \sqrt{nw}$,
an event which occurs with probability at least $1 - 2e^{-\lambda^2/4}$. 
Noting that $|\log d_{\vec \rho} - nMw| = M\left|\sum_i X_i \right| < M\lambda \sqrt{nw},$
we see that
$$
nMw - M \lambda \sqrt{nw} < \log d_{\vec \rho}
$$
and hence
$$
d_{\vec \rho} > \exp\left(nMw - M \lambda \sqrt{nw}\right).
$$
Given a constant $\beta > 0$, we choose $\lambda = \alpha \sqrt n$ where $\alpha = \sqrt w - \beta/(M\sqrt w),$
so that 
$$ 
d_{\vec \rho} > \exp\left(nMw - nM\left(\sqrt w - \beta/(M\sqrt w)\right)\sqrt{w}\right) 
= \exp \left(nMw - nM \left( w - \beta/M \right) \right) = \exp(\beta n),
$$
with probability at least $1 - 2e^{-\alpha^2n/4}.$ Clearly, we need $\lambda > 0$ to get a useful bound, 
and so we require that $\beta < Mw = \text{Exp}_\tau \log d_\tau.$
\end{proof}

\section{Strong Fourier Sampling on $G^n$}

In this section, we show that a stronger condition on the base group $G$ 
implies that even strong Fourier sampling cannot distinguish certain
subgroups of $G^n$. Roughly, the condition requires that the dimensionwise mass 
of the group algebra of $G$ consisting of representations that do \emph{not} satisfy 
(\ref{char-bound-condition}) is less than the expected dimension of 
an irreducible of $G$ chosen according to the Plancherel distribution. In Section
4 we will show that nonabelian simple groups satisfy this condition.

\subsection{A stronger condition on the base group}

Given an irreducible representation $\tau$ of a finite group, and an element $g$ of the group,
it is always the case that the \emph{normalized character} $\left| \chi_\tau(g)/d_\tau \right|$ of $\tau$,
evaluated at $g$, is at most equal to $1.$ Given a group $G$ and an involution $\mu \in G$,
let $\Delta$ denote the collection of irreducibles of $G$ for which the normalized character 
evaluated at $\mu$ is exactly equal to $1$, i.e.
\begin{equation}\label{delta-definition}
\Delta = \left \{ \tau \in \widehat G ~:~ \left|\chi_{\tau} (\mu)/d_{\tau}\right| = 1 \right \}.
\end{equation}
It is clear that $\Delta$ contains all of the one-dimensional irreducibles. We
wish to consider groups which satisfy the condition
\begin{equation}\label{G-condition}
\sum_{\tau \in \Delta} d_\tau^2 < e^{\text{Exp}_\tau\log d_\tau},
\end{equation}
where the expectation is taken over $\tau$ chosen according to $\mathcal P_G.$ Roughly stated,
we require that the dimension of the subspace of $\mathbb C[G]$ occupied by representations from
$\Delta$ is less than the expected dimension of an irreducible of $G$. We remark that
this condition implies $\sum_{\tau \in \Delta} d_\tau^2 < \max_{\tau \in \widehat G} d_\tau,$
and hence there exists at least one normalized character of $G$ which is less than one
at $\mu$. The following proposition shows that condition (\ref{G-condition}) is a strengthening of the 
requirement that $G$ simply contain an involution $\mu \notin Z(G).$

\begin{prop}
Let $G$ be a finite group. Then $g \in Z(G)$ if and only if
$|\chi_\tau(g)/d_\tau| = 1$ for every $\tau \in \widehat G.$
\end{prop}
\begin{proof}
First, if $g \in Z(G)$, then $\tau(g)$ commutes with $\tau(h)$
for all $\tau \in \widehat G$ and all $h \in G.$ By Schur's Lemma, $\tau(g)$ is a homothety $\lambda \identity_{d_\tau}$,
and hence $|\chi_\tau(g)| = |{\bf tr}\tau(g)| = d_\tau|\lambda| = d_\tau,$ for all $\tau$. On the other hand, 
suppose $|\chi_\tau(g)| = d_\tau$ for every $\tau$. Since the norm of the sum of $d_\tau$ complex numbers, each of norm one, only equals $d_\tau$ if they are all equal, we conclude that each $\tau(g)$ is a homothety, and thus commutes with $\tau(h)$ for every $h$. Since the regular 
representation is a direct sum of the $\tau$, it evaluated at $g$ must also commute with itself
evaluated at any $h$. As the regular representation is faithful, we have $g \in Z(G)$.
\end{proof}

\subsection{Normalized character bounds}

Let $G$ be a group and $\mu \in G$ an involution, such that condition (\ref{G-condition})
is satisfied. We now establish upper bounds for the normalized characters of 
$G^n$ evaluated at the conjugacy class $[\mu]^n = [(\mu, \dots, \mu)]$. Recall that a representation $\vec \sigma \in \widehat{G^n}$ 
is a tensor product $\vec \sigma = \sigma_1 \otimes \sigma_2 \otimes \cdots \otimes \sigma_n$ of $n$ factors from $\widehat G$, 
and that the character of such a representation is the product of the characters of 
its factors. The normalized character of $\vec \sigma$ then satisfies
$$
\left| \frac{\chi_{\vec \sigma}([\mu]^n)}{d_{\vec \sigma}} \right| = 
\prod_{i=1}^n \left| \frac{\chi_{\sigma_i}([\mu])}{d_{\sigma_i}} \right| =
\prod_{i:\sigma_i \notin \Delta} \left| \frac{\chi_{\sigma_i}([\mu])}{d_{\sigma_i}} \right|
\leq c^{-|\{\sigma_i \notin \Delta\}|} < 1,
$$
where
$$
c = \max \left\{ \left| \frac{\chi_\tau([\mu])}{d_\tau} \right| : \tau \in \widehat G \setminus \Delta \right \} ^{-1} > 1
$$
is a constant depending only on $G$. For some constant $0 < \epsilon < 1$ 
to be determined later, let $\Lambda$ denote
the set of irreducibles of $G^n$ whose tensor product
decomposition (into irreducibles of $G$) contains more than $1 - \epsilon$
elements of $\Delta.$ The normalized character of any irreducible $\vec \sigma \notin \Lambda$ 
is then exponentially small: 
\begin{equation}\label{c-bound}
\left|\frac{\chi_{\vec \sigma}\left([\mu]^n\right)}{d_{\vec \sigma}}\right| \leq c^{-\epsilon n} = e^{-\epsilon (\ln c) n}
\end{equation}
As part of the proof that strong Fourier sampling fails, we will need to show that the
variance term from Lemma~\ref{exp-var-lemma} is exponentially small. Equation (\ref{c-bound})
is a good bound for the normalized characters of representations outside $\Lambda$.
The rest of the representations comprise $L_{\vec \rho}$, the $\Lambda$-isotypic subspace of $\vec \rho \otimes \vec \rho^*$, 
where $\vec \rho \in \widehat{G^n}$. The multiplicity of any single $\vec \tau \in \widehat{G^n}$ in the 
direct sum decomposition of $\vec \rho \otimes \vec \rho^*$ can be bounded as follows:
$$
 \left \langle \chi_{\vec \tau}, \chi_{\vec \rho \otimes \vec \rho^*} \right \rangle
 = \left \langle \chi_{\vec \tau}, \chi_{\vec \rho} \cdot \chi_{\vec \rho}^* \right \rangle
 = \left \langle \chi_{\vec \rho} \chi_{\vec \tau}, \chi_{\vec \rho}^* \right \rangle \leq d_{\vec \tau}.
$$
Here we have used the fact that the last inner product is the multiplicity
of $\vec \rho^*$ in the representation $\vec \rho \otimes \vec \tau,$ which is clearly
at most $d_{\vec \rho} \cdot d_{\vec \tau} / d_{\vec \rho}.$ Given $\vec \tau$, 
let $\tau_1 \otimes \tau_2 \otimes \cdots \otimes \tau_n$ be the decomposition of $\tau$ into irreducibles 
of $G$, and let $s = \lceil \epsilon n \rceil.$ We can now give a bound 
for the dimension of $L_{\vec \rho}$ which does not depend on $\vec \rho$:
\begin{align*}
\dim L_{\vec \rho} 
& \leq \sum_{\vec \tau \in \Lambda} d_{\vec \tau}^2
  = \sum_{\vec \tau \in \Lambda} d_{\tau_1}^2 \cdots d_{\tau_n}^2
  \leq {n \choose s} \sum_{\substack{\rho_1, \dots,~\rho_s \in ~\widehat G\\ 
        \sigma_1, \dots,~\sigma_{n-s} \in ~\Delta}} d_{\rho_1}^2 \cdots d_{\rho_s}^2 \cdot d_{\sigma_1}^2 \cdots d_{\sigma_{n-s}}^2\\
& = {n \choose s} \sum_{\rho_1, \dots, \rho_s \in \widehat G} d_{\rho_1}^2 \cdots d_{\rho_s}^2
       ~\cdot \sum_{\sigma_1, \dots, \sigma_{n-s} \in \Delta} d_{\sigma_1}^2 \cdots d_{\sigma_{n-s}}^2\\
& = {n \choose s} \left(\sum_{\rho \in \widehat G} d_\rho^2\right)^s
       \cdot \left(\sum_{\sigma \in \Delta} d_\sigma^2\right)^{n-s}
 = {n \choose s} |G|^s \cdot \left(\sum_{\sigma \in \Delta} d_\sigma^2\right)^{n-s}.
\end{align*}
If we let $\delta$ denote the dimensionwise fraction of the group algebra of $G$ consisting 
of representations from $\Delta$, we then have
\begin{equation}\label{L-bound}
\dim L_{\vec \rho}  \leq {n \choose s} |G|^s \left(\delta |G| \right)^{n-s}
= {n \choose s} \delta^{n-s}|G|^n \approx \left(\frac{e}{\epsilon}\right)^{\epsilon n} \cdot \delta^{(1-\epsilon)n}|G|^n
\end{equation}

\subsection{Strong Fourier sampling fails}

We are now ready to prove that strong Fourier sampling cannot distinguish certain subgroups
of $G^n$.

\begin{theorem} \label{strong-samp-thm}
Let $G$ be a group containing an involution $\mu$ such that condition
(\ref{G-condition}) is satisfied. Let $B = \{{\bf b}\}$ be a frame with weights $\{a_{\bf b}\}$ satisfying the 
completeness condition (\ref{completeness}) for an irreducible representation
$\vec \rho \in \widehat{G^n}$. Given the hidden subgroup $H_m = \{1, m\}$ where $m$ is chosen
uniformly at random from the conjugacy class $[\mu]^n$, let $P({\bf b})$ be the
probability that we observe the vector ${\bf b}$ conditioned on having
observed the representation name $\vec \rho$, and let $N$ be the natural distribution on $B$.
Then there are positive constants $\kappa$ and $\eta$ such that for sufficiently large $n$,
$$
\| P - N \|_1 < e^{-\kappa n}
$$
with probability at least $1-e^{-\eta n}$ in $m$ and $\vec \rho.$
\end{theorem}

\begin{proof}
Throughout the proof, we will maintain the notation established in the preceding sections. For simplicity,
we assume here that $B$ is an orthonormal basis. The proof is easily modified for the case where
$B$ is a frame, just as in \cite{MRS05}. Recall that
\begin{equation*}
P({\bf b}) = P_{H_m, \vec \rho} ({\bf b}) = \frac{\left \|\Pi_m {\bf b}\right \|^2}{{\bf rk}~\Pi_m}
\qquad
\text{and}
\qquad
N({\bf b}) = P_{\{1\}, \vec \rho} = \frac{1}{d_{\vec \rho}}
\end{equation*}
The condition (\ref{G-condition})  provides for the existence of a positive constant $\gamma$ such that
\begin{equation}\label{gamma-bound}
\delta |G| = \sum_{\tau \in \Delta} d_\tau^2 < e^{-\gamma} e^{\text{Exp}_\tau \left[ \log d_\tau \right]}.
\end{equation}

In order to control the variance of $\left \|\Pi_m {\bf b}\right \|^2$, we must control the number of basis 
vectors which project significantly into $L$, the $\Lambda$-isotypic subspace 
of $\vec \rho \otimes \vec \rho^*.$ By (\ref{L-bound}),
$$
\dim L \leq \left[\left(\frac{e}{\epsilon}\right)^{\epsilon} \delta^{(1-\epsilon)}|G|\right]^n.
$$
Let $\Pi_L$ denote the projector onto $L$, and let $B_L$ be the collection of basis vectors 
${\bf b} \in B$ that violate the bound
\begin{equation}\label{prop-L}
\left \| \Pi_L({\bf b} \otimes {\bf b}^*) \right \|^2 < e^{- \gamma n/2},
\end{equation} 
and hence may project significantly into the subspace $L$.
As the ${\bf b} \otimes {\bf b}$ are orthogonal, we can bound $|B_L|$ by counting dimensions:
$$
|B_L| \leq e^{\gamma n/2} \dim L 
\leq \left[ \left(\frac{e}{\epsilon}\right)^{\epsilon} e^{\gamma / 2} \delta^{(1-\epsilon)}|G|\right]^n.
$$

Consider a particular ${\bf b} \notin B_L.$ We can control the variance of $\left \|\Pi_m {\bf b} \right \|^2$
via Lemma~\ref{exp-var-lemma}. Recall that, for $\vec \sigma \notin \Lambda,$ we have 
the bound (\ref{c-bound}). Hence
\begin{align*}
\text{Var}_m \left \| \Pi_m {\bf b} \right \|^2
  \leq &\frac{1}{4} \left[ \sum_{\vec \sigma \prec \vec \rho \otimes \vec \rho^*\text{, }\vec \sigma \in \Lambda} 
          \left \|\Pi^{\vec \rho}_{\vec \sigma}({\bf b} \otimes {\bf b}^*) \right \|^2
    + \sum_{\vec \sigma \prec \vec \rho \otimes \vec \rho^*\text{, }\vec \sigma \notin \Lambda} \frac{\chi_{\vec \sigma}(m)}{d_{\vec \sigma}} 
          \left \|\Pi^{\vec \rho}_{\vec \sigma}({\bf b} \otimes {\bf b}^*) \right \|^2 \right]  \\
  \leq &\frac{1}{4} \left[ e^{-\gamma n/2} + e^{-\epsilon (\ln c) n} \sum_{\vec \sigma \notin \Lambda} 
          \left \|\Pi^{\vec \rho}_{\vec \sigma}({\bf b} \otimes {\bf b}^*) \right \|^2 \right]
  \leq \frac{1}{4}\left[e^{-\gamma n/2} + e^{-\epsilon (\ln c) n}\right]
  \leq \frac{1}{2}~e^{-4an}
\end{align*}
where we have let $a = \min\{\gamma/2, \epsilon(\ln c)\}/4 > 0$ for simplicity.
The above shows that $\left \| \Pi_m {\bf b} \right \|^2$ will be very close to its expectation with overwhelming probability. Indeed, by Chebyshev's inequality, 
the probability that
\begin{equation}\label{goodvect}
\left|\left\|\Pi_m{\bf b}\right\|^2 - \text{Exp}_m\left \|\Pi_m{\bf b}\right \|^2\right| 
\leq e^{-an}
\end{equation}
is at least $1 - (e^{-4an}/2)/(e^{-an})^2 = 1 - e^{-2an}/2.$ Let $B_{\text{bad}}$ denote the vectors in 
$B$ which violate (\ref{goodvect}), so that
$$
\text{Exp}_m \left|B_{\text{bad}}\right| \leq  \frac{1}{2}~e^{-2an} |B| = \frac{1}{2}~e^{-2an} d_{\vec \rho}.
$$
We now condition on three events:
\begin{itemize}
\item $E_1$: $\left|B_{\text{bad}}\right| \leq e^{-a n} d_{\vec \rho}$
\item $E_2$: $d_{\vec \rho} \geq e^{\beta n}$ for some $0 < \beta < \text{Exp}_{\tau \in \widehat G}\log d_\tau$
\item $E_3$: $\vec \rho \notin \Lambda. $
\end{itemize}
By Markov's inequality, $\text{Pr}[E_1] \geq 1 - e^{- an}$. By 
Lemma~\ref{big-irrep-lemma}, $\text{Pr}[E_2] \geq 1 - 2e^{-\alpha^2n/4}$ where $\alpha > 0$
is as defined in the Lemma. For the probability of $E_3$'s occurrence, recall that the probability of choosing a 
particular $\vec \rho$ according to weak sampling is never more 
than twice the probability of selecting it according to $\mathcal P_{G^n}$, which is in turn 
equal to the probability of selecting each of its $n$ tensor product factors independently 
according to $\mathcal P_G$. The probability of choosing an element in $\Delta$ from $\widehat G$ 
in this way is at most $\delta < 1$. Hence $E_3$ occurs with probability at least
$$
1 - 2{n \choose \epsilon n}\delta^{(1-\epsilon)n} \approx 1 - 2\left(\frac{e}{\epsilon}\right)^{\epsilon n} \delta^{(1-\epsilon)n}.
$$
Since $\epsilon^\epsilon \to 1$ as $\epsilon \to 0$, we can choose $\epsilon > 0$
small enough so that $E_2$ occurs with overwhelming probability.

We now separate the total variation distance between $P$ and $N$ as follows:
$$
\| P - N \|_1 = \sum_{{\bf b} \notin B_L \cup B_{bad}} |P({\bf b}) - N({\bf b})| 
   + \sum_{{\bf b} \in B_L \cup B_{bad}} |P({\bf b}) - N({\bf b})|.
$$
First, consider the sum over the nice vectors ${\bf b} \notin B_L \cup B_{bad}$, which
satisfy conditions (\ref{prop-L}) and (\ref{goodvect}). By (\ref{expect}) and (\ref{rankbound}), 
\begin{align*}
\sum_{{\bf b} \notin B_L \cup B_{bad}} |P({\bf b}) - N({\bf b})|
& = \sum_{{\bf b} \notin B_L \cup B_{bad}} \left|\frac{\left\|\Pi_m{\bf b}\right\|^2}{{\bf rk}~\Pi_m}
 - \frac{1}{d_{\vec \rho}}\right|
 = \sum_{{\bf b} \notin B_L \cup B_{bad}} \left|\frac{\left\|\Pi_m{\bf b}\right\|^2}{{\bf rk}~\Pi_m}
 - \frac{\text{Exp}_m\left \|\Pi_m{\bf b}\right \|^2}{{\bf rk}~\Pi_m}\right| \\
& \leq d_{\vec \rho} \cdot \left|\frac{e^{-an}}{{\bf rk}~\Pi_m}\right|
  \leq \frac{e^{-an}d_{\vec \rho}}{d_{\vec \rho}\left(1 + \chi_{\vec \rho}(m)/d_{\vec \rho}\right)/2} \leq \frac{2e^{-an}}{1-e^{-\epsilon (\ln c) n}} \leq 4e^{-an}
\end{align*}
for sufficiently large $n$.
It follows that $P(B_L \cup B_{\text{bad}})$ is at most $\left| B_L \cup B_{\text{bad}}\right|/d_{\vec \rho} 
+ 4 e^{-an}$. Since we conditioned on $E_1$, we already have $\left|B_{\text{bad}}\right| \leq e^{-an} 
d_{\vec \rho}.$ We wish to achieve a similar bound for $B_L.$ By (\ref{gamma-bound}),
$$
e^{\gamma/2}\delta |G| < e^{-\gamma/2} e^{\text{Exp}_\tau\left[ \log d_\tau \right]}.
$$
Since $\epsilon^\epsilon \to 1$ as $\epsilon \to 0$, we can choose $\epsilon$ 
smaller if necessary, so that
$$
e^{\gamma/2}\left(\frac{e}{\epsilon}\right)^\epsilon\delta^{1-\epsilon} |G| < e^{-\gamma/2} e^{\text{Exp}_\tau\left[ \log d_\tau \right]}.
$$
Since we can choose $\beta$ arbitrarily close to $\text{Exp}_\tau\left[ \log d_\tau \right]$, we can achieve
$$
|B_L| \leq \left[e^{\gamma/2}\left(\frac{e}{\epsilon}\right)^\epsilon\delta^{1-\epsilon} |G|\right]^n 
< e^{-\gamma n/2} e^{\beta n} \leq e^{-\gamma n/2} d_{\vec \rho}
$$
as desired. Thus $\left|B_L \cup B_{\text{bad}} \right| \leq e^{-\gamma n/2}d_{\vec \rho} + e^{-an}d_{\vec \rho}$. The second sum is then
\begin{align*}
\sum_{{\bf b} \in B_L \cup B_{bad}} |P({\bf b}) - N({\bf b})| &\leq P(B_L \cup B_{\text{bad}}) + \frac{\left|B_L \cup B_{\text{bad}}\right|}{d_{\vec \rho}}
   \leq \frac{2\left| B_L \cup B_{\text{bad}}\right|}{d_{\vec \rho}} + 4 e^{-an}\\
 &\leq 2e^{-\gamma n/2} + 6 e^{-an}.
\end{align*}
Combining the sum, we have
$$
\|P - N \|_1 \leq 2e^{-\gamma n/2} + 10 e^{-an},
$$
with probability at least $\text{Pr}\left[E_1 \land E_2 \right] \geq 1 - e^{-an} - 4e^{-\alpha^2n/4}$ where $\alpha$ is the constant
from Lemma \ref{big-irrep-lemma}.
\end{proof}

\section{Examples}

\subsection{Simple base groups}

Let $G$ be nonabelian and simple, i.e. having only trivial normal subgroups. 
By the Feit-Thompson theorem, $G$ has even order, and thus contains a nontrivial involution $\mu$. 
The nontrivial representations of $G$ are faithful, since their kernels are normal and hence 
must equal the identity subgroup. Thus there are no nontrivial 
one-dimensional representations of $G$, as they would in fact be isomorphisms of $G$ 
with subgroups of the abelian multiplicative group of complex numbers of norm one. 
Suppose that $\rho$ is an irreducible of $G$ with $d_\rho > 1$ and  $|\chi_\rho(\mu)| = d_\rho$. 
Then $\rho(\mu)$ is a homothety, and hence it commutes with $\rho(g)$ for every $g.$
By the faithfulness of $\rho$, $\mu \in Z(G) = \{1\}$, a contradiction. We conclude that 
only the trivial representation has normalized character equal to one. 
Clearly, $G$ satisfies (\ref{G-condition}), and hence one register is insufficient
to distinguish subgroups of $G^n.$

\subsection{Solvable groups}

Consider the wreath product $\ZZ_2 \wr G \cong \ZZ_2 \ltimes (G \times G)$
of a finite group $G$ with $\ZZ_2$. The action of $\ZZ_2$ in the semidirect
product is the 'flip-flop' action on the two copies of $G$. For example,
$$
(1, (a, b))\circ (0, (c, d)) = (1 + 0, (a, b)\cdot 1(c, d)) = (1, (a, b) \cdot (d, c)) = (1, (ad, bc))
$$
for every $a, b, c, d \in G.$
The irreducible representations of $\ZZ_2 \wr G$ are characterized as follows.
The irreducibles of $G \times G$ are tensor products
$\rho \otimes \sigma$, where $\rho$ and $\sigma$ are $G$-irreducibles.
Each of these induces up to a representation 
$$
\theta_{\rho, \sigma} = \mathbf{Ind}_{_{G \times G}}^{^{\ZZ_2 \wr G}} \rho \otimes \sigma
$$
of $\ZZ_2 \wr G.$ The space of this representation is a direct sum of two
copies of the space of $\rho \otimes \sigma$. The action of $\theta_{\rho, \sigma}(1, (a, b))$
swaps these two subspaces, and thus has trace zero. The action of $\theta_{\rho, \sigma}(0, (a, b))$
decomposes into a $\rho(a) \otimes \sigma(b)$ action on the first subspace, and
a $\rho(b) \otimes \sigma(a)$ action on the second subspace. We thus have
$$
\chi_{_{\theta_{\rho,\sigma}}}(x, (a, b)) =
\begin{cases}
\chi_\rho(a) \chi_\sigma(b) + \chi_\rho(b) \chi_\sigma(b) & \text{~~if x = 0} \\
0 & \text{~~if x = 1}
\end{cases}
$$
To calculate the decompositions of the $\theta_{\rho, \sigma}$, we compute
the relevant inner products:
\begin{align*}
\bra \chi_{_{\theta_{\rho_1,\sigma_1}}}, \chi_{_{\theta_{\rho_2,\sigma_2}}}\ket
&= \frac{1}{2|G|^2}\sum_{a, b \in G} \left(\chi_{\rho_1}(a)\chi_{\sigma_1}(b) + \chi_{\rho_1}(b)\chi_{\sigma_1}(a)\right)
                 \cdot \left(\chi_{\rho_2}^*(a)\chi_{\sigma_2}^*(b) + \chi_{\rho_2}^*(b)\chi_{\sigma_2}^*(a)\right)\\
& = \bra \chi_{\rho_1}, \chi_{\rho_2}\ket \bra \chi_{\sigma_1}, \chi_{\sigma_2}\ket + \bra \chi_{\rho_1}, \chi_{\sigma_2}\ket \bra \chi_{\sigma_1}, \chi_{\rho_2}\ket
\end{align*}
Hence, if $\rho \neq \sigma$, then $\theta_{\rho,\sigma}$ is irreducible and 
$\theta_{\rho, \sigma} \cong \theta_{\sigma, \rho}.$ On the other hand,
$\theta_{\rho, \rho}$ decomposes into two irreducibles, each with multiplicity
one and dimension $d_\rho^2.$ We will denote these irreducibles by $\theta_{\rho, \rho}^+$
and $\theta_{\rho, \rho}^-,$ corresponding to their actions on the space of $\rho \otimes \rho$,
which are described by $(1, (a, b))\cdot u \otimes v \mapsto bv \otimes au$ and
$(1, (a, b))\cdot u \otimes v \mapsto -bv \otimes au$, respectively. Observe that
\begin{align*}
\left|\chi_{_{\theta_{\rho,\rho}^+}}(x, (a, b))\right| 
= \left|\chi_{_{\theta_{\rho,\rho}^-}}(x, (a, b))\right|
& = \sum_{u, v \in \rho} \bra u \otimes v, (x, (a, b)) \cdot u \otimes v \ket \\
& = \begin{cases}
\sum_{u, v \in \rho} \bra u, au \ket \bra v, bv \ket = \chi_\rho(a)\chi_\rho(b) = \chi_\rho(ab) & \text{~if x = 0} \\
\sum_{u, v \in \rho} \bra u, bv \ket \bra v, au \ket = \sum_{u, v \in \rho} \rho(b)_{uv}\chi_\rho(a)_{vu} = \chi_\rho(ab) & \text{~if x = 1.} \\
\end{cases}
\end{align*}
Hence, for these representations, the normalized character is always at most $1/d_\rho.$

Now consider the group $\ZZ_2 \wr D_k,$ and suppose for simplicity that $k$ is odd. The dihedral
group $D_k$ contains an involution $F$ (the 'flip'). It has two one-dimensional and 
$(k-1)/2$ two-dimensional ireducible representations. By the above, the group $\ZZ_2 \wr D_k$ will thus have:
\begin{itemize}
\item $4$ one-dimensional irreducibles $\theta_{\rho, \rho}^\pm$, 
      where $\rho$ is one-dimensional in $\widehat{D_k},$
\item $1$ two-dimensional irreducible $\theta_{\rho, \sigma}$, 
      where $\rho$ and $\sigma$ are one-dimensional,
\item $k-1$ four-dimensional irreducibles $\theta_{\rho, \sigma}$, where $\rho$ is one-dimensional
      and $\sigma$ is two-dimensional,
\item $k-1$ four-dimensional irreducibles $\theta_{\rho, \rho}^\pm$, where $\rho$ is two-dimensional, and
\item $\frac{1}{2}\frac{(k-1)}{2}\frac{(k-3)}{2}$  eight-dimensional irreducibles $\theta_{\rho, \sigma}$, where 
$\rho$ and $\sigma$ are two-dimensional.
\end{itemize}
We know that the irreducibles of the form $\theta_{\rho, \rho}^+$ and
$\theta_{\rho, \rho}^-$ have normalized characters equal to $1/d_\rho$. Observe that if we choose
the involution $\mu = (1, (F, F))$, then the irreducibles $\theta_{\rho, \sigma}$
have normalized character $0$ at $\mu.$ Hence the only 'bad' irreducibles of $\ZZ_2 \wr D_k$
are one dimensional, and thus this group satisfies condition (\ref{G-condition}) for sufficiently large $n$. 
Our results then imply that groups of the form
$(\ZZ_2 \wr D_k)^n$ defy strong Fourier sampling on one register. Recall that $D_n$
is solvable, and observe that $D_n \times D_n$ is an index two normal subgroup of $\ZZ_2 \ltimes (D_n \times D_n)$.
Hence $\ZZ_2 \wr D_n$ is solvable, and since we can write
$$
\ZZ_2 \wr D_k \lhd (\ZZ_2 \wr D_k)^2 \lhd (\ZZ_2 \wr D_k)^3 \lhd \cdots \lhd (\ZZ_2 \wr D_k)^n,
$$
so are the groups $(\ZZ_2 \wr D_k)^n$. We remark that the factors in the normal series
for these groups are of size constant in $n$.

\subsection{Nilpotent groups}

Consider the group $\ZZ_2^{\wr 2} := \ZZ_2 \wr \ZZ_2 \cong \ZZ_2 \ltimes (\ZZ_2 \times \ZZ_2).$ By the above analysis,
$\ZZ_2^{\wr 2}$ has four one-dimensional irreducible representations, and one which is two-dimensional. Repeating
the construction, we see that $\ZZ_2^{\wr 3} = \ZZ_2 \wr \ZZ_2^{\wr 2}$ will have:
\begin{itemize}
\item $8$ one-dimensional irreducibles $\theta_{\rho, \rho}^\pm$, 
      where $\rho$ is one-dimensional in $\widehat{\ZZ_2^{\wr 2}},$
\item $6$ two-dimensional irreducibles $\theta_{\rho, \sigma}$, 
      where $\rho$ and $\sigma$ are one-dimensional,
\item $4$ four-dimensional irreducibles $\theta_{\rho, \sigma}$, where $\rho$ is one-dimensional
      and $\sigma$ is two-dimensional, and
\item $2$ four-dimensional irreducibles $\theta_{\rho, \rho}^\pm$, where $\rho$ is two-dimensional.
\end{itemize}
In general, it is not hard to see that each iteration of this construction, yielding groups $\ZZ_2^{\wr k}$,
doubles the number of one-dimensional irreducibles, while the rest increase at least quadratically. We choose the involution 
$$
\mu = \left(1, \left(0_{\ZZ_2^{\wr k-1}}, 0_{\ZZ_2^{\wr k-1}}\right)\right) \in \ZZ_2^{\wr k},
$$ 
where $0_G$ denotes the identity element of $G$.
As in the previous section, the normalized characters of all non-one-dimensional irreducibles,
evaluated at $\mu$, will then be bounded below one. Hence, for sufficiently
large $k$, the groups $\ZZ_2^{\wr k}$ satisfy our condition (\ref{G-condition}),
and thus strong sampling cannot succeed on groups of the form $\left(\ZZ_2^{\wr k}\right)^n.$
Clearly, these are $2$-groups, and hence nilpotent.

\end{document}